\pdfoutput=1
\documentclass[a4paper,twoside,12pt]{article}
\usepackage{arxiv}
\usepackage{subfiles}

\usepackage{iftex}
\ifPDFTeX
  \usepackage[T1]{fontenc}
  \usepackage{lmodern}       
  \usepackage[scaled]{helvet} 
\else
  \usepackage{fontspec}       
\fi

\usepackage{graphicx}   
\usepackage{import}     
\usepackage{float}      
\usepackage{wrapfig}    
\usepackage{lscape}     

\usepackage{algorithm}
\usepackage{algpseudocode} 

\usepackage{listings}

\usepackage{caption}
\DeclareCaptionLabelSeparator{nn}{ \textbar\ } 
\usepackage{subcaption}
\usepackage{alphalph}   
\renewcommand\thesubfigure{\roman{subfigure}}
\makeatletter
\renewcommand\p@subfigure{\thefigure:}
\makeatother

\usepackage{sidecap}
\usepackage{xpatch}

\AtBeginDocument{\raggedbottom}
\usepackage{booktabs, multirow} 
\usepackage{ltablex}  
\keepXColumns         
\usepackage{array}    

\usepackage{amsmath}
\usepackage{amssymb}
\usepackage{siunitx}
\usepackage[compact,raggedright]{titlesec} 
\usepackage{enumitem}  
\usepackage{hyperref} 
\hypersetup{
  colorlinks=true,
  linkcolor=blue,
  citecolor=blue,
  urlcolor=blue
}

\usepackage[nameinlink,capitalise,noabbrev]{cleveref} 
\crefname{figure}{Fig.}{Figs.}
\Crefname{figure}{Fig.}{Figs.}
\crefname{subfigure}{Fig.}{Figs.}
\Crefname{subfigure}{Fig.}{Figs.}

\usepackage[acronym,toc]{glossaries} 
\usepackage{afterpage}  
\usepackage{comment}    

\usepackage[numbers,sort&compress,super]{natbib}
\newcommand{\autocite}[1]{\cite{#1}}
\newcommand{\textcite}[1]{\cite{#1}}

\usepackage[auth-sc]{authblk}

\graphicspath{{imgs/}{imgs/main/}{imgs/sup/}}
\DeclareGraphicsExtensions{.pdf,.png,.jpg,.jpeg}

\newcommand{\beginsupplement}{%
  \setcounter{table}{0}%
  \renewcommand{\thetable}{S\arabic{table}}%
  \setcounter{figure}{0}%
  \renewcommand{\thefigure}{S\arabic{figure}}%
}

\title{DNATokenizer: A GPU-First Byte-to-Identifier Tokenizer for High-Throughput DNA Language Models}

\date{\vspace{-5ex}}

\newcommand{\emailthanks}[1]{\hspace{0.15em}\thanks{\href{mailto:#1}{#1}}}

\author[1]{Eliatan Niktab\emailthanks{eli.niktab@anu.edu.au}}
\author[1]{Hardip Patel}
\affil[1]{\raggedright National Centre for Indigenous Genomics,
Australian National University, Canberra, ACT, Australia}

\newcommand{\DNATok}{\textsc{DNATok}}

\begin{document}

\clearpage

\maketitle

\begin{abstract}
Tokenization sits at the boundary between high-throughput genomic input and GPU compute, posing a challenge for both algorithmic design choices and systems bottlenecks. Inherently, overlapping $k$-mer tokenization introduces information leakage under masked language modeling (MLM) and can degrade downstream accuracy; single-nucleotide tokenization avoids leakage and preserves per-base fidelity but explodes sequence length for attention-based architectures. Non-overlapping $k$-mers and byte-pair encoding (BPE) provide compression and avoid leakage, at the cost of boundary sensitivity or reduced interpretability. Empirically, the choice of tokenization interacts strongly with model architecture and task requirements. At the system level, however, standard string tokenizers and host-bound vocabulary lookups dominate wall-clock time once inputs reach billions of bases, regardless of the chosen algorithmic tokenization.

This work presents \DNATok, a high-performance GPU-first tokenization system that replaces general-purpose string processing with byte lookup table (LUT)-based identifier streaming and an overlapped host-to-device (H2D)/compute pipeline utilizing pinned memory and architectural parallelism. \DNATok\ is vocabulary-agnostic; it accelerates single-nucleotide, non-overlapping $k$-mer, or BPE tokenization and integrates as a drop-in systems layer beneath genomic foundation models. On the benchmark system described in Table~\ref{tab:system_config}, \DNATok\ achieves $84$--$95\times$ higher encode throughput than optimized Hugging Face baselines and up to $1.9\times$ higher H2D throughput. End-to-end streaming throughput reaches $1.27$--$1.84\times 10^{8}$ tokens/s depending on configuration, effectively removing tokenization as a bottleneck for production-scale training and inference.
\end{abstract}

\section{Introduction}
Deep learning models increasingly rely on efficient input pipelines to realize their highly parallelized potential. For DNA sequences, tokenization sits at the critical boundary between high-throughput I/O and GPU compute. Modern genomic models, including Nucleotide Transformer,\autocite{dallatorre2025nt} HyenaDNA,\autocite{nguyen2023hyenadna} Evo2,\autocite{nguyen2024evo} Caduceus,\autocite{schiff2024caduceus} and GPN\autocite{benegas2023gpn} demonstrate that tokenization choice fundamentally shapes model performance, in terms of hardware utilization, training efficiency, and generalization across species and tasks.

\subsection{Algorithmic tokenization: the leakage problem}
Early genomic language models employed overlapping $k$-mer tokenization with stride 1, where each token is a substring of length $k$ and adjacent tokens share $k{-}1$ bases.\autocite{ji2021dnabert} This approach captures local sequence motifs and provides richer immediate context, but introduces two critical problems. First, it produces redundant token streams of length $L{-}k{+}1$ for an input of length $L$, increasing memory and compute costs. Second, and more seriously, overlapping $k$-mers cause \emph{information leakage} under masked language modeling: when predicting a masked token, the model can infer much of the answer from adjacent unmasked tokens that share $k{-}1$ bases with the target, degrading sample efficiency and potentially harming downstream accuracy.\autocite{zhou2023dnabert2}

BPE-based genomic tokenizers address this leakage problem by replacing overlapping $k$-mers with learned subword vocabularies.\autocite{zhou2023dnabert2,sennrich2016bpe} BPE constructs a vocabulary of variable-length tokens from frequent genomic substrings, providing substantial input compression, eliminating the stride-1 overlap, and thereby avoiding MLM leakage.\autocite{zhou2023dnabert2}

Non-overlapping (blocked) $k$-mers compress input length in proportion to $k$ and avoid MLM leakage by eliminating shared bases between adjacent tokens. However, they introduce boundary sensitivity: small shifts, insertions, or deletions can substantially change tokenization, potentially missing motifs that cross token boundaries. BPE and other data-driven subword schemes provide variable-length tokens that adapt to frequent genomic patterns, offering strong compression and few out-of-vocabulary tokens, but may lack direct biological interpretability and can exhibit domain-transfer challenges across species.\autocite{zhou2023dnabert2,lindsey2025tokenizers}

Single-nucleotide (character) tokenization preserves per-base resolution and is robust to small insertions and deletions, making it ideal for tasks sensitive to single-nucleotide polymorphisms and for architectures that efficiently handle long contexts (for example, state-space models like HyenaDNA and Evo2).\autocite{nguyen2023hyenadna,nguyen2024evo} However, character tokenization produces very long token sequences that stress attention-based models at long context lengths.

Table~\ref{tab:genomic-tokenization} summarizes major tokenization schemes and their empirical trade-offs as synthesized in recent reviews and comparative studies.\autocite{lindsey2025tokenizers,zhou2023dnabert2}

\begin{table}[htbp]
\centering
\caption{Genomic tokenization schemes and their trade-offs for long-context DNA language models.}
\label{tab:genomic-tokenization}
\begin{tabular}{>{\raggedright\arraybackslash}p{0.18\textwidth} >{\raggedright\arraybackslash}p{0.26\textwidth} >{\raggedright\arraybackslash}p{0.26\textwidth} >{\raggedright\arraybackslash}p{0.18\textwidth}}
\toprule
\textbf{Scheme} & \textbf{Benefits} & \textbf{Drawbacks} & \textbf{Benchmarked models} \\
\midrule
Single-nucleotide (character) &
Per-base resolution; robust to single-nucleotide variants and small indels; no token-boundary artefacts. &
Sequences as long as the input DNA; high compute cost for attention at long context lengths. &
HyenaDNA,\autocite{nguyen2023hyenadna} Evo2\autocite{nguyen2024evo} \\
\addlinespace[0.4em]
Overlapping $k$-mers &
Enriches local context; captures short motifs. &
Adjacent tokens share $k-1$ bases, causing MLM leakage and redundancy; now generally discouraged.\autocite{zhou2023dnabert2} &
Not benchmarked \\
\addlinespace[0.4em]
Non-overlapping $k$-mers &
Reduces input length in proportion to $k$; avoids MLM leakage. &
Boundary sensitivity: small shifts or indels change the tokenization frame and can split motifs. &
Nucleotide Transformer\autocite{dallatorre2025nt} \\
\addlinespace[0.4em]
BPE / data-driven subwords &
Strong compression; compact vocabulary; avoids MLM leakage. &
Reduced biological interpretability; domain-transfer challenges across genomes and assays. &
Not benchmarked \\
\bottomrule
\end{tabular}
\end{table}

\subsection{The systems gap}
Genomic models provide limited systems-level approaches to overcome tokenization as a throughput bottleneck in production workloads. The focus remains primarily on token design, model architectures, and downstream benchmarks (e.g., GenomicBenchmarks\autocite{gresova2023genomicbenchmarks}); explicit, quantitative studies of tokenization throughput, pinned host memory, host-device concurrency, or GPU-resident tokenization are sparse.\autocite{zhou2023dnabert2,lindsey2025tokenizers} Recent genomic transformer work highlights model-side accelerations such as FlashAttention and Attention with Linear Biases (ALiBi),\autocite{dao2022flashattention,press2021alibi} but does not detail input pipeline optimizations. Broader LLM systems surveys acknowledge tokenization as a pipeline step affecting latency and throughput, and discuss concurrent CUDA streams and overlapping transfers with compute, but reproducible measurements specific to tokenizer throughput in genomics remain limited.\autocite{cudaGuide,ptPin,olcfConcurrency}

NVIDIA programming guides and systems papers on deep learning emphasize pinned memory reuse, asynchronous transfers, and compute-communication overlap as prerequisites for high-throughput GPU applications.\autocite{cudaGuide,ptPin,olcfConcurrency} However, these principles have not been systematically applied or quantified for genomic tokenization. Existing work usually treats tokenization as a solved preprocessing step, ignoring the substantial serialization overhead imposed by CPU-bound string processing at the terabyte scale. \DNATok\ bridges this gap by providing a dedicated, reproducible pipeline optimized for fixed-alphabet genomic inputs, compatible with recent foundation models including Evo2, Nucleotide Transformer, and HyenaDNA.

In practical genomic embedding stacks, popular string tokenizers from NLP toolkits, most notably the Hugging Face tokenizers library,\autocite{wolf2020transformers} incur per-character branching, dynamic allocations, and host-side vocabulary lookups that are unnecessary for DNA's fixed four-letter alphabet. For long contexts and large batches, tokenization becomes both CPU-bound and H2D-bound, dominating wall-clock time even when downstream models are highly optimized. This systems bottleneck is agnostic to whether the user employs single-nucleotide, $k$-mer, or BPE tokenization at the algorithmic level.

\subsection{Contributions}
This paper presents \DNATok, a tokenization system engineered to eliminate string-processing overheads for DNA at scale while remaining compatible with the benchmarked tokenizers for Nucleotide Transformer (NTv2/NTv3), HyenaDNA, and Evo2. \DNATok's design is orthogonal to algorithmic tokenization choice; it accelerates whatever vocabulary the user selects and provides three complementary pathways:
\begin{enumerate}
  \item \textbf{Tokenizer-agnostic systems layer.} A drop-in framework that accelerates single-nucleotide, non-overlapping $k$-mer, or BPE tokenization by converting the process into memory movement and fixed-size gather operations, independent of the underlying vocabulary choice.
  \item \textbf{Optimized encoding paths.} We provide three complementary strategies: (a) an \emph{IDs path} using vectorized CPU mapping and pinned staging; (b) a \emph{bytes path} utilizing device-side 256-entry LUTs; and (c) a \emph{pipelined streaming path} that hides H2D latency by overlapping transfers with embedding computation via CUDA streams.
  \item \textbf{Rigorous empirical validation.} We evaluate \DNATok\ across 222 benchmark configurations, including synthetic microbenchmarks with variable batch sizes and sequence lengths, and real-world integration with NTv3, HyenaDNA, and Evo2 models. This validation confirms consistent speedups of $20$--$112\times$ across diverse model architectures.
\end{enumerate}

\section{Design}

\DNATok\ is a GPU-accelerated tokenization system that replaces string-based tokenization with lookup-table-based identifier streaming. The core insight is that genomic tokenization, whether single-nucleotide, $k$-mer, or BPE, reduces to mapping ASCII characters to integer identifiers, an operation well-suited to vectorization and GPU parallelism. Our design provides multiple encoding strategies optimized for different hardware bottlenecks, automatic discovery of tokenizer metadata, and seamless compatibility with existing genomic foundation models.

Let $B$ denote batch size, $T$ sequence length, and $D$ embedding dimension. Assume an embedder exposing
\[
  \texttt{embed\_tokens} : \mathbb{Z}^{B\times T} \rightarrow \mathbb{R}^{B\times T\times D}.
\]

\DNATok\ focuses on accelerating the conversion of raw DNA strings into the integer identifier space required by this function.

\subsection{Discovery and validation}

\DNATok\ probes the upstream tokenizer and embedding stack to retrieve: the padding identifier, valid DNA and ambiguity identifiers, the 256-entry ASCII$\rightarrow$identifier LUT, and any fixed token length requirement. A self-test on \texttt{ACGTNacgtn} verifies consistency between the upstream tokenizer (for example, Hugging Face tokenizers for NTv2/NTv3 or custom vocabularies for HyenaDNA and Evo2) and the discovered LUT. Failures trigger a fallback path that calls the upstream string tokenizer directly to preserve correctness.

\subsection{Encoding paths}

\DNATok\ provides three complementary encoding strategies that address different performance bottlenecks. Users select strategies via the high-level \texttt{embed\_from\_strings()} API, which automatically chooses optimal paths based on hardware capabilities or accepts manual overrides for specific optimization goals.

\paragraph{IDs path (CPU staging).}
The IDs path performs vectorized ASCII$\rightarrow$identifier conversion on CPU using NumPy into persistent pinned host buffers, then launches non-blocking H2D copies as 32- or 64-bit integers. \DNATok\ provides two implementations: a legacy \texttt{encode\_batch\_to\_ids()} that allocates new tensors per call, and an optimized \texttt{encode\_batch\_to\_ids\_staging()} that reuses persistent pinned buffers across invocations. The staging variant supports both int32 and int64 output formats, reducing H2D traffic when smaller integer widths suffice. This path is the simplest drop-in acceleration for existing tokenizers, converting tokenization into memory movement plus a gather operation. On-device type promotion from int32 to int64 (when needed) is nearly free, making 32-bit staging preferable when PCIe bandwidth is the bottleneck.

\paragraph{Bytes path (on-device LUT).}
The bytes path stages raw 8-bit ASCII in pinned memory, transfers to the device, then applies a device-resident 256-entry LUT to obtain identifiers. This reduces H2D traffic relative to 32- or 64-bit identifiers, which is particularly beneficial when the interconnect is the dominant bottleneck. The fixed-size LUT fits easily in L1 cache or shared memory. Unlike the IDs path, the bytes path performs a single H2D transfer followed by entirely on-device processing (LUT application, optional left-padding, micro-batching), eliminating the need for ping-pong buffers or stream coordination.

\begin{algorithm}[H]
\caption{\texttt{DiscoverIDsPath} --- enable IDs path and build ASCII$\to$ID LUT}
\begin{algorithmic}[1]
\Require Embedder $E$ with method $\texttt{embed\_tokens}(\cdot)$
        and optional tokenizer
\Ensure  Flag \texttt{use\_ids\_path},
         lookup table \texttt{ascii\_lut[0..255]},
         IDs \texttt{id\_pad}, \texttt{id\_N},
         optional fixed length \texttt{token\_len}
\Statex
\State Let $f \gets E.\texttt{embed\_tokens}$
\If{$f$ is not callable}
  \State \texttt{use\_ids\_path} $\gets$ \textbf{False}
  \State \Return
\EndIf
\State $pad \gets$ padding ID discovered from $E$ (pad token fields or pad token string; default $0$)
\State $tok \gets$ tokenizer associated with $E$ (underlying tokenizer if wrapped)
\If{$tok$ is \textbf{None}}
  \Comment{fixed DNA fallback}
  \State Initialise $\texttt{ascii\_lut}$ of length $256$ with value $0$
  \State Define
    $$
      \text{A/a} \mapsto 1,\;
      \text{C/c} \mapsto 2,\;
      \text{G/g} \mapsto 3,\;
      \text{T/t} \mapsto 4,\;
      \text{N/n} \mapsto 0
    $$
  \For{\textbf{each} DNA character $c \in \{\text{A,C,G,T,N,a,c,g,t,n}\}$}
    \State $\texttt{ascii\_lut}[\mathrm{ord}(c)] \gets \text{ID}(c)$
  \EndFor
  \State $\texttt{id\_pad} \gets pad,\quad \texttt{id\_N} \gets 0$
  \State \texttt{use\_ids\_path} $\gets$ \textbf{True}
  \State \Return
\EndIf
\Statex
\State Try detect K-mer structure (length $k$, stride $k$)
\If{K-mer structure detected}
  \State Build $\texttt{kmer\_lut}[0..5^k-1]$ and $\texttt{base5\_lut}[0..255]$
  \State \texttt{use\_ids\_path} $\gets$ \textbf{True}
  \State \Return
\EndIf
\Comment{Standard tokenizer-based path (Char/BPE)}
\State For each $c \in \{\text{A,C,G,T,N,a,c,g,t,n}\}$, let $dna\_ids[c]$ be
       the token ID discovered for $c$ (using tokenizer APIs), if any
\State Make $dna\_ids$ case-insensitive for A,C,G,T and ensure
       $dna\_ids[\text{``N''}]$ and $dna\_ids[\text{``n''}]$ exist;
       if no ID for ``N''/``n'' is found, set both to $pad$
\State $n\_id \gets dna\_ids[\text{``N''}]$
\State Initialise $\texttt{ascii\_lut}$ of length $256$ with value $n\_id$
\For{$b = 0,\dots,255$}
  \State Try to obtain a token ID representing byte $b$ from $tok$
        (via a byte-level token or a single-character token)
  \If{such an ID exists}
    \State $\texttt{ascii\_lut}[b] \gets \text{that ID}$
  \EndIf
\EndFor
\For{\textbf{each} $(c, id)$ in $dna\_ids$}
  \State $\texttt{ascii\_lut}[\mathrm{ord}(c)] \gets id$
\EndFor
\State $\texttt{id\_pad} \gets pad,\quad \texttt{id\_N} \gets n\_id$
\State $\texttt{token\_len} \gets$ first positive value among
       known maximum sequence length fields of $E$ (if any)
\State \texttt{use\_ids\_path} $\gets$ \textbf{True}
\end{algorithmic}
\end{algorithm}

\newpage

\begin{algorithm}[H]
\caption{\texttt{EmbedFromStrings} --- streaming DNA embeddings via IDs path}
\begin{algorithmic}[1]
\Require Equal-length strings $S = \{s_i\}_{i=1}^B$, each of length $T$,
         embedding batch cap $B_{\mathrm{emb}}$,
         target device \texttt{acc},
         path $\in\{\texttt{"ids"},\texttt{"bytes"},\texttt{"auto"}\}$
\Require IDs path already discovered:
         \texttt{use\_ids\_path} = \textbf{True},
         \texttt{ascii\_lut}, \texttt{id\_pad},
         optional \texttt{token\_len},
         tokens-per-call budget $M$
\Ensure Streamed embeddings on \texttt{acc}
\Statex
\If{$path = \texttt{"auto"}$}
  \State $path \gets \texttt{"bytes"}$ if \texttt{acc} is a CUDA device
         with GPU available, else $\texttt{"ids"}$
\EndIf
\Statex
\If{$path = \texttt{"bytes"}$}
  \Comment{device-side ASCII$\to$ID mapping}
  \State Pack $S$ into a byte matrix $\mathrm{Bytes}[B \times T]$
  \State Transfer $\mathrm{Bytes}$ to \texttt{acc} as $\mathrm{Bytes}_{\texttt{acc}}$
  \If{$k$-mer mode enabled}
    \State $\mathrm{IDs}_{\texttt{acc}} \gets \texttt{MapKmer}(\mathrm{Bytes}_{\texttt{acc}}, \texttt{kmer\_lut})$
  \Else
    \State $\mathrm{IDs}_{\texttt{acc}} \gets
           \texttt{ascii\_lut}[\mathrm{Bytes}_{\texttt{acc}}]$
  \EndIf
  \If{$\texttt{token\_len}$ is defined and $\texttt{token\_len} > T$}
    \State Left-pad each row of $\mathrm{IDs}_{\texttt{acc}}$ with \texttt{id\_pad},
           right-aligning the original $T$ positions
    \State $T' \gets \texttt{token\_len}$
  \Else
    \State $T' \gets T$
  \EndIf
  \State $b \gets \min\!\bigl(B_{\mathrm{emb}},\;
       \max(1,\lfloor M / T' \rfloor)\bigr)$ \Comment{micro-batch size}
  \For{$lo \gets 0$ \textbf{to} $B-1$ \textbf{step} $b$}
    \State $hi \gets \min(B,\, lo + b)$
    \State Call $\texttt{embed\_tokens}(\mathrm{IDs}_{\texttt{acc}}[lo:hi])$
           on \texttt{acc} and emit the resulting activations
  \EndFor
  \State \Return
\EndIf
\Statex
\Comment{Host-side IDs path}
\State Map $S$ to IDs on the host using \texttt{ascii\_lut},
       obtaining $\mathrm{IDs}_{\mathrm{host}}[B \times T]$
\If{$\texttt{token\_len}$ is defined and $\texttt{token\_len} > T$}
  \State Left-pad each row of $\mathrm{IDs}_{\mathrm{host}}$ with \texttt{id\_pad},
         right-aligning the original $T$ positions; set $T' \gets \texttt{token\_len}$
\Else
  \State $T' \gets T$
\EndIf
\State $b \gets \min\!\bigl(B_{\mathrm{emb}},\;
       \max(1,\lfloor M / T' \rfloor)\bigr)$
\For{$lo \gets 0$ \textbf{to} $B-1$ \textbf{step} $b$}
  \State $hi \gets \min(B,\, lo + b)$
  \State Transfer $\mathrm{IDs}_{\mathrm{host}}[lo:hi]$ to \texttt{acc}
  \State Call $\texttt{embed\_tokens}$ on this slice and emit activations
\EndFor
\end{algorithmic}
\end{algorithm}

\paragraph{Pipelined streaming.}
Pipelined streaming applies exclusively to the IDs path, overlapping H2D transfers with embedding compute using separate CUDA streams. The bytes path does not use pipelined streaming. The pipelined path pre-allocates device buffers. When using int32 staging, four buffers are allocated (two int32 for H2D staging, two int64 for embedder calls); when using int64 direct mode, two int64 buffers suffice. CUDA streams and events coordinate the overlap.\autocite{cudaGuide}
\begin{itemize}
  \item \textit{Streaming baseline} (\DNATok) uses sequential H2D copy followed by embedding compute per micro-batch (no overlap).
  \item \textit{Streaming pipelined} (\DNATok) uses dual streams and ping-pong device buffers so that H2D transfer for micro-batch $n{+}1$ is overlapped with embedding compute for micro-batch $n$.
\end{itemize}
Both variants keep tokenizer logic unchanged; the difference lies purely in how data movement is scheduled relative to compute. The impact of pipelining is workload-dependent and is quantified in the Results section.

\subsection{Robustness and compatibility}

\begin{description}
    \item[Memory management and automatic micro-batch sizing.] \DNATok\ automatically calculates micro-batch size as $b = \min(B_{\mathrm{emb}}, \max(1, \lfloor M / T \rfloor))$ where $M$ is the token budget (default 1,048,576 tokens per call). This prevents 32-bit index overflow in embedding operations. To handle out-of-memory or overflow despite this sizing, we detect failures (\texttt{"canuse32bitindexmath"}, \texttt{"out of memory"}, \texttt{"conv1d index"}), halve micro-batch size, and retry. Persistent buffers (\texttt{\_staging\_ids\_cpu}, \texttt{\_staging\_bytes\_cpu}, device ping-pong buffers) are reused across calls to avoid repeated allocations.
    
    \item[Cascading fallback hierarchy.] Three levels of graceful degradation ensure correctness: (1) pipelined streaming falls back to baseline streaming on mid-pipeline errors without duplicating outputs, (2) baseline streaming retries with halved batch sizes on OOM/overflow, (3) discovery failure triggers fallback to the upstream string tokenizer. Errors in byte parsing or LUT application trigger safe fallback to the IDs path or upstream tokenizer.
    
    \item[Compatibility validation.] \DNATok\ integrates with the Hugging Face-style tokenizers used by NTv2/NTv3, HyenaDNA, and Evo2. Optional self-tests compare token IDs on random genomic subsequences (with special-token handling where applicable) to confirm tokenization equivalence during setup.
\end{description}

\subsection{User interface}

Users interact with \DNATok\ primarily through the \texttt{embed\_from\_strings} API call, which handles the complete pipeline from strings to embeddings. The \texttt{path} parameter controls strategy selection:
\begin{description}
  \item[\texttt{"auto"} (default):] Selects \texttt{"bytes"} on CUDA devices, \texttt{"ids"} otherwise
  \item[\texttt{"ids"}:] Forces IDs path with optional pipelined streaming (controlled by \texttt{overlap\_h2d\_compute} initialization flag)
  \item[\texttt{"bytes"}:] Forces bytes path with on-device LUT application
\end{description}
This abstraction allows optimization for different hardware configurations (PCIe vs. NVLink, bandwidth-limited vs. compute-limited) without modifying application code.

\section{Results}

\subsection{Experimental Setup}
\textbf{Hardware/Software.} See Table~\ref{tab:system_config}. \textbf{Workloads.} Standard, large-batch, and long-sequence configurations follow the settings reported in Table~\ref{tab:benchmark_results}. \textbf{Baselines.} Hugging Face tokenizer (batched interface).\autocite{wolf2020transformers} \textbf{\textnormal{\DNATok} variants.} IDs path (64-bit), IDs path (32-bit + on-device cast), streaming baseline, streaming pipelined.

We evaluate \DNATok\ against Hugging Face's standard tokenizer implementation across three key dimensions: raw encoding throughput, host-to-device transfer efficiency, and end-to-end streaming performance. All experiments use a fixed nucleotide vocabulary, and the standard configuration aligns with the benchmark table unless otherwise noted.

\subsection{Encoding Throughput}

\begin{figure}[H]
\centering

\begingroup
  \captionsetup{font=small,skip=4pt}
  \captionsetup[subfigure]{labelformat=parens,justification=centering}
  \renewcommand\thesubfigure{\roman{subfigure}}

  \begin{subfigure}[b]{0.78\textwidth}
    \centering
    \includegraphics[width=\linewidth,height=0.32\textheight,keepaspectratio]{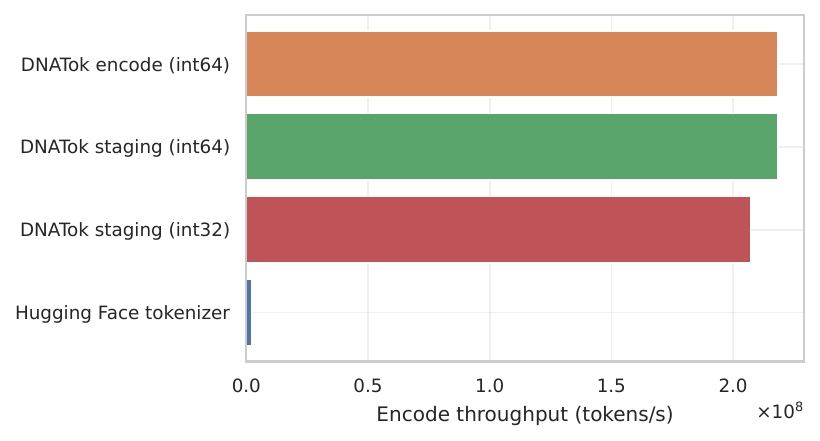}
    \caption{}\label{fig:encode-standard-panel}
  \end{subfigure}

  \vspace{0.05cm}

  \caption[Encode throughput—standard configuration]{\textbf{Encode throughput (tokens per second) by method for the standard configuration (batch size $B=4096$, sequence length $T=512$).} DNATok variants outperform the Hugging Face tokenizer baseline by $90$--$95\times$, with staging i64 and encode i64 tied and staging i32 slightly lower; see Table~\ref{tab:benchmark_results} for exact values.}
  \label{fig:encode-standard}
\endgroup
\end{figure}

\DNATok's GPU-native encoding eliminates the CPU bottleneck, delivering $84$--$95\times$ speedups over the optimized Hugging Face baseline in microbenchmarks (Figure~\ref{fig:encode-standard}; Table~\ref{tab:benchmark_results}). In the standard configuration, 64-bit encoding (both staging and direct) saturates the device's integer throughput, significantly outperforming the 32-bit path which incurs casting overhead. This ordering persists across large-batch and long-sequence regimes, confirming that minimizing host-side ops is the dominant factor.

\begin{figure}[H]
\centering

\begingroup
  \captionsetup{font=small,skip=4pt}
  \captionsetup[subfigure]{labelformat=parens,justification=centering}
  \renewcommand\thesubfigure{\roman{subfigure}}

  \begin{subfigure}[b]{0.78\textwidth}
    \centering
    \includegraphics[width=\linewidth,height=0.32\textheight,keepaspectratio]{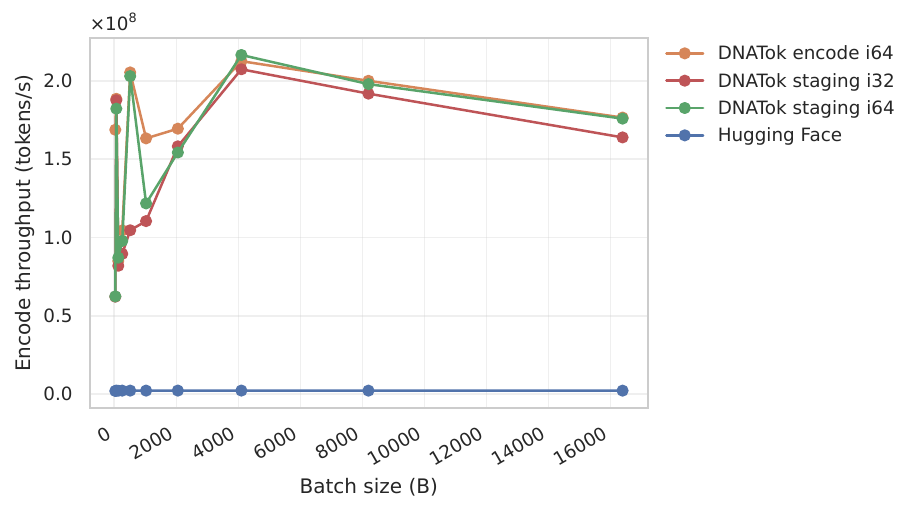}
    \caption{}\label{fig:batch-sweep-encode-panel}
  \end{subfigure}

  \vspace{0.05cm}

  \caption[Encode throughput vs batch size]{\textbf{Encode throughput versus batch size sweep.} DNATok throughput rises with batch size and then plateaus, while tokenizer baselines remain relatively flat; trends align with the benchmark summary.}
  \label{fig:batch-sweep-encode}
\endgroup
\end{figure}

\begin{figure}[H]
\centering

\begingroup
  \captionsetup{font=small,skip=4pt}
  \captionsetup[subfigure]{labelformat=parens,justification=centering}
  \renewcommand\thesubfigure{\roman{subfigure}}

  \begin{subfigure}[b]{0.78\textwidth}
    \centering
    \includegraphics[width=\linewidth,height=0.32\textheight,keepaspectratio]{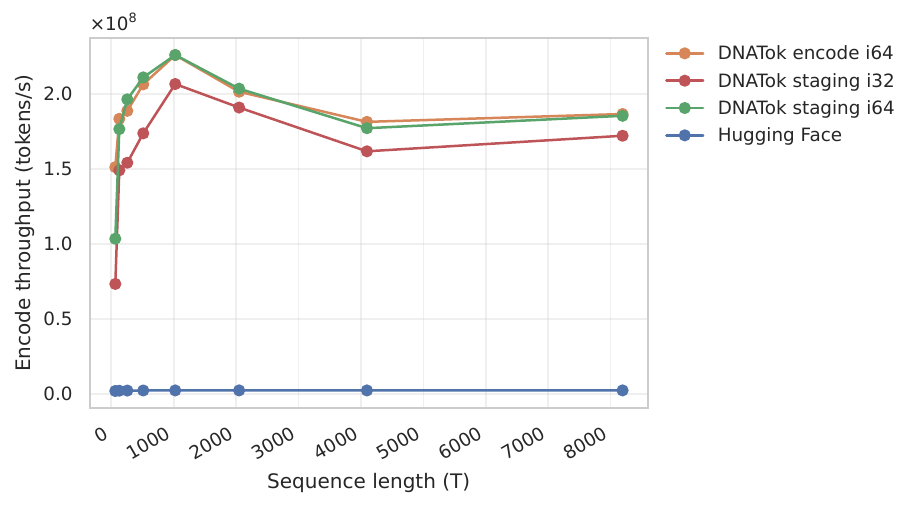}
    \caption{}\label{fig:length-sweep-encode-panel}
  \end{subfigure}

  \vspace{0.05cm}

  \caption[Encode throughput vs sequence length]{\textbf{Encode throughput versus sequence length sweep.} DNATok throughput varies with sequence length and plateaus across mid-to-long lengths, while Hugging Face tokenizer baselines remain comparatively flat.}
  \label{fig:length-sweep-encode}
\endgroup
\end{figure}

\begin{figure}[H]
\centering

\begingroup
  \captionsetup{font=small,skip=4pt}
  \captionsetup[subfigure]{labelformat=parens,justification=centering}
  \renewcommand\thesubfigure{\roman{subfigure}}

  \begin{subfigure}[b]{0.78\textwidth}
    \centering
    \includegraphics[width=\linewidth,height=0.32\textheight,keepaspectratio]{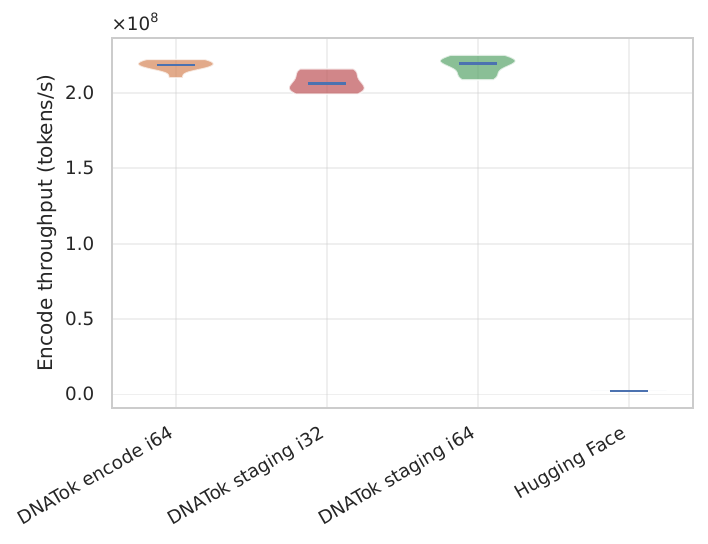}
    \caption{}\label{fig:encode-violin-panel}
  \end{subfigure}

  \vspace{0.05cm}

  \caption[Encode throughput distribution—violin plot]{\textbf{Distribution of encode throughput across repetitions (violin plot).} DNATok variants show higher medians and modest variance compared to tokenizer baselines, indicating stable measurements across trials.}
  \label{fig:violin-encode}
\endgroup
\end{figure}

Performance scaling across batch sizes reveals the expected saturation regimes (Figure~\ref{fig:batch-sweep-encode}), while sequence-length sweeps show similar trends (Figure~\ref{fig:length-sweep-encode}). Trial-to-trial variability remains moderate (Figure~\ref{fig:violin-encode}), supporting reproducibility across runs.

\subsection{Host-to-Device Transfer Efficiency}

\begin{figure}[H]
\centering

\begingroup
  \captionsetup{font=small,skip=4pt}
  \captionsetup[subfigure]{labelformat=parens,justification=centering}
  \renewcommand\thesubfigure{\roman{subfigure}}

  \begin{subfigure}[b]{0.78\textwidth}
    \centering
    \includegraphics[width=\linewidth,height=0.32\textheight,keepaspectratio]{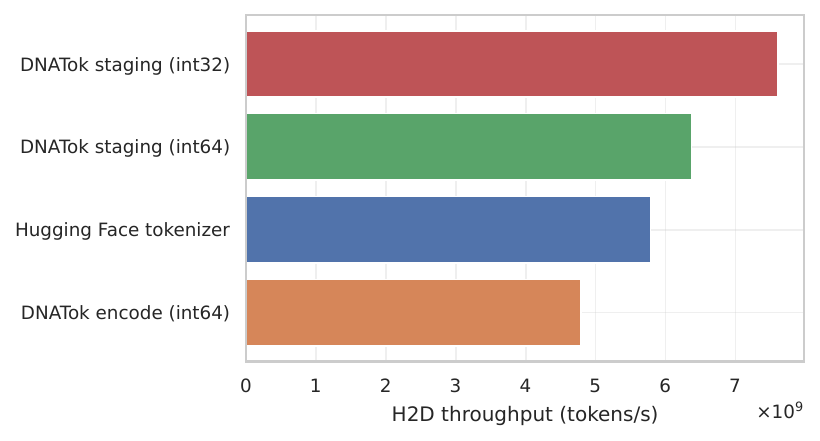}
    \caption{}\label{fig:h2d-standard-panel}
  \end{subfigure}

  \vspace{0.05cm}

  \caption[Host-to-Device throughput—standard configuration]{\textbf{Host-to-Device throughput (tokens per second) by method for the standard configuration.} Int32 staging delivers the highest H2D throughput by reducing transfer volume; staging i64 is slightly above the tokenizer baseline, while encode i64 is lower. Widening to 64-bit is performed on device.}
  \label{fig:h2d-standard}
\endgroup
\end{figure}

Host-to-device (H2D) transfer efficiency is maximized by minimizing data volume. The int32 staging path achieves the highest H2D throughput ($1.3$--$1.9\times$ baseline) by halving the transfer payload compared to 64-bit paths (Figure~\ref{fig:h2d-standard}). This optimization is impactful because the baseline Hugging Face implementation is bottlenecked by serialized CPU-side memory copies rather than PCIe bandwidth. The 32-bit path effectively doubles the available transfer budget, a critical optimization for PCIe-constrained systems.

\subsection{End-to-End Streaming Performance}

\begin{figure}[H]
\centering

\begingroup
  \captionsetup{font=small,skip=4pt}
  \captionsetup[subfigure]{labelformat=parens,justification=centering}
  \renewcommand\thesubfigure{\roman{subfigure}}

  \begin{subfigure}[b]{0.78\textwidth}
    \centering
    \includegraphics[width=\linewidth,height=0.32\textheight,keepaspectratio]{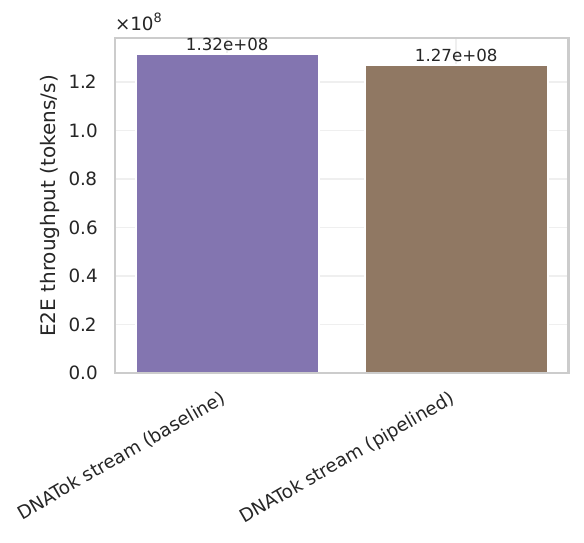}
    \caption{}\label{fig:e2e-standard-panel}
  \end{subfigure}

  \vspace{0.05cm}

  \caption[End-to-end streaming—baseline vs pipelined]{\textbf{End-to-end throughput comparing baseline and pipelined streaming.} In the standard configuration, pipelining slightly underperforms baseline streaming; see Table~\ref{tab:benchmark_results} for measured values.}
  \label{fig:e2e-standard}
\endgroup
\end{figure}

Overlapping compute and transfer via pipelined streaming yields workload-dependent outcomes (Figure~\ref{fig:e2e-standard}). In the standard configuration, pipelining slightly underperforms baseline streaming, while it improves throughput in the large-batch and long-sequence configurations (Table~\ref{tab:benchmark_results}).

\begin{figure}[H]
\centering

\begingroup
  \captionsetup{font=small,skip=4pt}
  \captionsetup[subfigure]{labelformat=parens,justification=centering}
  \renewcommand\thesubfigure{\roman{subfigure}}

  \begin{subfigure}[b]{0.78\textwidth}
    \centering
    \includegraphics[width=\linewidth,height=0.32\textheight,keepaspectratio]{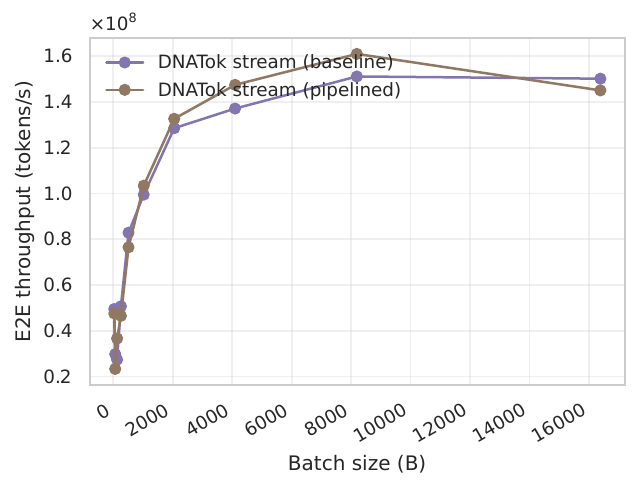}
    \caption{}\label{fig:batch-sweep-e2e-panel}
  \end{subfigure}

  \vspace{0.05cm}

  \caption[End-to-end throughput vs batch size]{\textbf{End-to-end throughput versus batch size.} Pipelined streaming shows mixed behavior: it underperforms at the smallest batches, exceeds baseline for several mid-to-large batches, and converges near parity at the largest batches.}
  \label{fig:batch-sweep-e2e}
\endgroup
\end{figure}

\begin{figure}[H]
\centering

\begingroup
  \captionsetup{font=small,skip=4pt}
  \captionsetup[subfigure]{labelformat=parens,justification=centering}
  \renewcommand\thesubfigure{\roman{subfigure}}

  \begin{subfigure}[b]{0.78\textwidth}
    \centering
    \includegraphics[width=\linewidth,height=0.32\textheight,keepaspectratio]{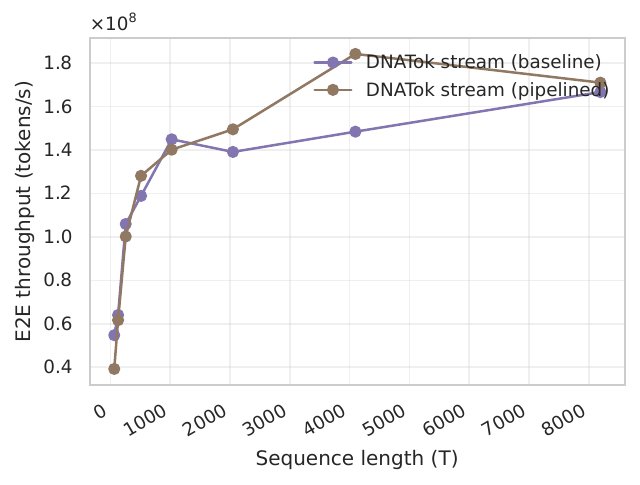}
    \caption{}\label{fig:length-sweep-e2e-panel}
  \end{subfigure}

  \vspace{0.05cm}

  \caption[End-to-end throughput vs sequence length]{\textbf{End-to-end throughput versus sequence length.} Pipelined streaming lags at very short lengths and improves for longer sequences, with mixed gains across the sweep.}
  \label{fig:length-sweep-e2e}
\endgroup
\end{figure}

The magnitude and direction of pipelining gains depends on the compute-to-transfer ratio (Figures~\ref{fig:batch-sweep-e2e}, \ref{fig:length-sweep-e2e}). Batch sweeps show mixed behavior with a mid-range crossover: pipelining underperforms at small batches, exceeds baseline for some mid-to-large batches, and converges at the largest batches. Length sweeps show pipelining lagging at very short lengths, improving at longer lengths, and fluctuating around parity depending on sequence length.

\subsection{Ablation Study}

\begin{figure}[H]
\centering

\begingroup
  \captionsetup{font=small,skip=4pt}
  \captionsetup[subfigure]{labelformat=parens,justification=centering}
  \renewcommand\thesubfigure{\roman{subfigure}}

  \begin{subfigure}[b]{0.78\textwidth}
    \centering
    \includegraphics[width=\linewidth,height=0.32\textheight,keepaspectratio]{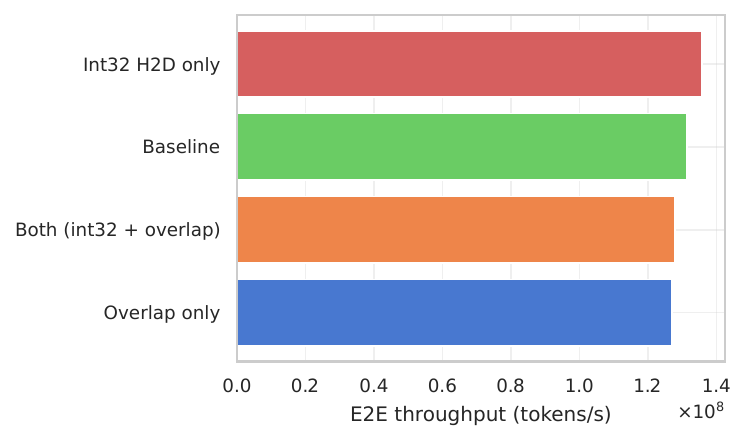}
    \caption{}\label{fig:ablation-e2e-panel}
  \end{subfigure}

  \vspace{0.05cm}

  \caption[Ablation—int32 transfers and overlap]{\textbf{Ablation study quantifying individual contributions to end-to-end throughput.} Int32-only is marginally above baseline, while overlap-only and int32+overlap are slightly below; see Table~\ref{tab:ablation_study} for measured values.}
  \label{fig:ablation-e2e}
\endgroup
\end{figure}

To isolate optimization contributions, we compare four configurations (Table~\ref{tab:ablation_study}; Figure~\ref{fig:ablation-e2e}). In the standard regime ($B=4096, T=512$), simply enabling overlap or int32 transfers yields negligible gains ($<1.05\times$) or slight regressions because the compute kernel is too fast to hide transfer latency. This negative result clarifies that \DNATok's primary speedup comes from the vectorised CPU map and pinned memory staging (the "IDs path"), rather than CUDA stream overlap, which only becomes beneficial when the compute-to-transfer ratio increases (e.g., larger batches).

\subsection{Real-Model Benchmarks}

\begin{figure}[H]
\centering

\begingroup
  \captionsetup{font=small,skip=4pt}
  \captionsetup[subfigure]{labelformat=parens,justification=centering}
  \renewcommand\thesubfigure{\roman{subfigure}}

  \begin{subfigure}[b]{0.85\textwidth}
    \centering
    \includegraphics[width=\linewidth,height=0.4\textheight,keepaspectratio]{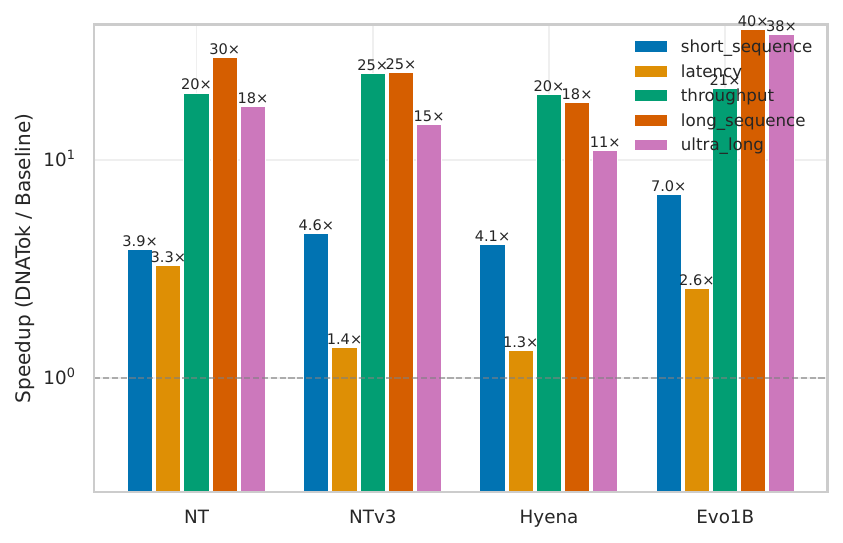}
    \caption{}\label{fig:real-speedup-grouped-panel}
  \end{subfigure}

  \vspace{0.05cm}

  \caption[Real-model speedups—grouped]{\textbf{Real-model speedups (DNATok vs. baseline) by scenario for NTv2/NTv3, HyenaDNA, and Evo2-1B.} Speedups are largest in throughput/long-sequence runs (roughly $20$--$40\times$) with smaller gains in latency and short-sequence runs; see Figure~\ref{fig:real-speedup-heatmap} for scenario-wise detail.}
  \label{fig:real-speedup-grouped}
\endgroup
\end{figure}

\begin{figure}[H]
\centering

\begingroup
  \captionsetup{font=small,skip=4pt}
  \captionsetup[subfigure]{labelformat=parens,justification=centering}
  \renewcommand\thesubfigure{\roman{subfigure}}

  \begin{subfigure}[b]{0.7\textwidth}
    \centering
    \includegraphics[width=\linewidth,height=0.32\textheight,keepaspectratio]{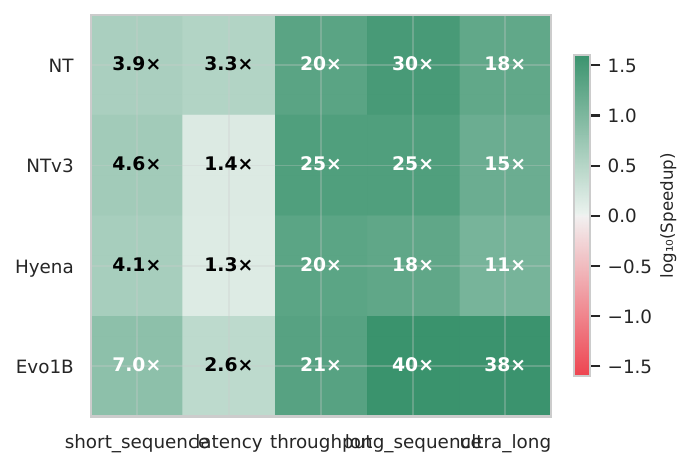}
    \caption{}\label{fig:real-speedup-heatmap-panel}
  \end{subfigure}

  \vspace{0.05cm}

  \caption[Real-model speedups—heatmap]{\textbf{Heatmap of real-model speedups (log$_{10}$ scale).} Entries show DNATok speedup relative to baseline for NTv2/NTv3, HyenaDNA, and Evo2-1B across short-sequence, latency, throughput, long-sequence, and ultra-long scenarios.}
  \label{fig:real-speedup-heatmap}
\endgroup
\end{figure}

\begin{figure}[H]
\centering

\begingroup
  \captionsetup{font=small,skip=4pt}
  \captionsetup[subfigure]{labelformat=parens,justification=centering}
  \renewcommand\thesubfigure{\roman{subfigure}}

  \begin{subfigure}[b]{0.9\textwidth}
    \centering
    \includegraphics[width=\linewidth,height=0.5\textheight,keepaspectratio]{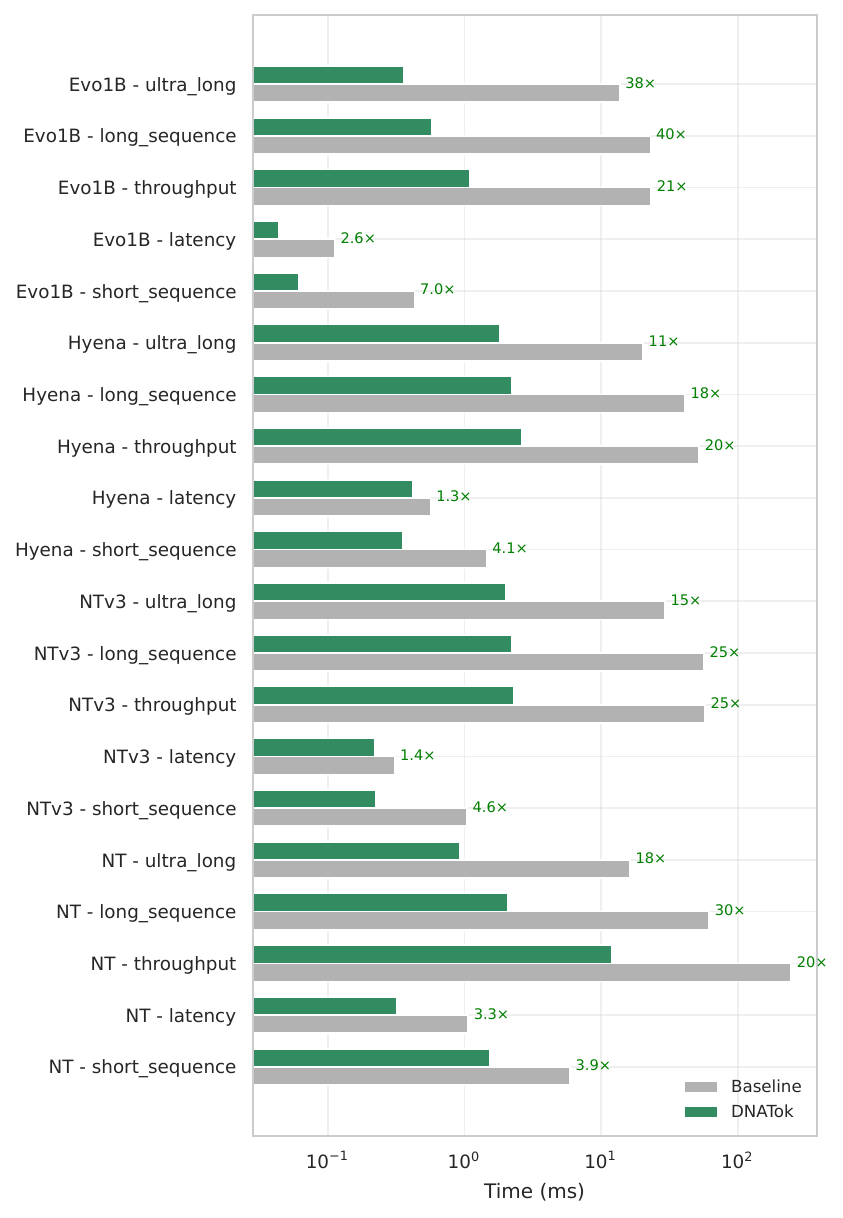}
    \caption{}\label{fig:real-latency-panel}
  \end{subfigure}

  \vspace{0.05cm}

  \caption[Real-model latency/throughput comparison]{\textbf{Latency and throughput across models and scenarios (log time scale).} Bars compare baseline vs. \DNATok\ for NTv2/NTv3, HyenaDNA, and Evo2-1B across short-sequence, latency-focused, throughput-focused, long-sequence, and ultra-long settings; annotated speedups match the grouped results.}
  \label{fig:real-latency}
\endgroup
\end{figure}

\begin{figure}[H]
\centering

\begingroup
  \captionsetup{font=small,skip=4pt}
  \captionsetup[subfigure]{labelformat=parens,justification=centering}
  \renewcommand\thesubfigure{\roman{subfigure}}

  \begin{subfigure}[b]{0.95\textwidth}
    \centering
    \includegraphics[width=\linewidth,height=0.42\textheight,keepaspectratio]{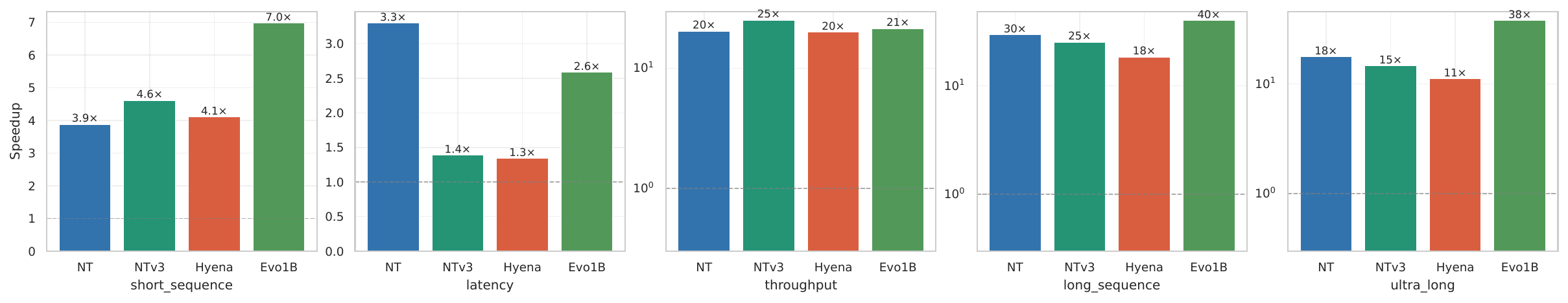}
    \caption{}\label{fig:real-scenario-panel}
  \end{subfigure}

  \vspace{0.05cm}

  \caption[Real-model speedups—by scenario]{\textbf{Speedups by scenario (short sequence, latency, throughput, long sequence, ultra-long) for NTv2/NTv3, HyenaDNA, and Evo2-1B.} Values shown correspond to the grouped benchmark outputs.}
  \label{fig:real-scenario}
\endgroup
\end{figure}

This study benchmarks end-to-end tokenization and embedding throughput for NTv2 (NTv2\_500M), NTv3 variants (8M/100M/650M), HyenaDNA (tiny\_1k/small\_32k/medium\_160k), and Evo2-1B under short-sequence, latency-focused, throughput-focused, long-sequence, and ultra-long scenarios (\Cref{fig:real-speedup-grouped,fig:real-speedup-heatmap,fig:real-latency,fig:real-scenario}). The grouped results show short-sequence gains of $3.88$--$6.98\times$, latency gains of $1.34$--$3.30\times$, throughput gains of $19.96$--$25.13\times$, long-sequence gains of $18.37$--$40.04\times$, and ultra-long gains of $11.09$--$37.81\times$; the per-model breakdown is provided in \texttt{results/benchmarks/real\_model\_benchmark.csv}.

\subsection{Summary of Performance Gains}

Across all metrics, \DNATok\ demonstrates:
\begin{itemize}
    \item \textbf{Encoding}: $84$--$95\times$ speedups over HF native baselines in the microbench configurations (Table~\ref{tab:benchmark_results})
    \item \textbf{H2D transfer}: int32 staging improves effective bandwidth up to $1.9\times$, while 64-bit paths remain near baseline (Figure~\ref{fig:h2d-standard})
    \item \textbf{E2E streaming}: pipelined streaming trails baseline in the standard configuration but improves throughput in the large-batch and long-sequence configurations (Figures~\ref{fig:e2e-standard}, \ref{fig:batch-sweep-e2e}, \ref{fig:length-sweep-e2e})
    \item \textbf{Reproducibility}: stable medians across trials with moderate variance (Figure~\ref{fig:violin-encode})
    \item \textbf{Scalability}: clear saturation regimes across batch and length sweeps (Figures~\ref{fig:batch-sweep-encode}, \ref{fig:length-sweep-encode})
\end{itemize}

These improvements reduce tokenization time in the benchmarked workloads, shifting more of the end-to-end budget to downstream preprocessing and model execution for the tested configurations.

\section{Discussion}

\subsection{Systems Optimization Orthogonal to Algorithmic Choices}

\DNATok\ does not change the underlying tokenization scheme—it is agnostic to whether users employ single-nucleotide, non-overlapping $k$-mer, or BPE tokenization. Those algorithmic decisions remain driven by downstream task requirements, model architecture, and masked language modeling (MLM) leakage considerations as established in the literature.\autocite{zhou2023dnabert2,lindsey2025tokenizers} Instead, \DNATok\ targets the systems bottleneck common to all tokenization schemes by replacing general-purpose string processing with lookup-table-based identifier streaming and overlapping host-to-device transfers with embedding compute.

This orthogonality is critical for adoption: algorithmic debates about optimal tokenization strategies for genomic foundation models continue,\autocite{lindsey2025tokenizers} and different models make different choices based on their architectural constraints and target tasks. By operating purely at the systems layer, \DNATok\ allows researchers to explore tokenization algorithms without being constrained by implementation performance. A vocabulary change requires only rebuilding the lookup table; the GPU kernels, memory management, and streaming infrastructure remain unchanged.

Measured speedups for encoding and H2D transfers cover character-level tokenization (HyenaDNA, Evo2) and non-overlapping $k$-mer tokenization (Nucleotide Transformer) in our benchmarks; we do not report BPE benchmarks here. The only requirement is that the vocabulary fits in GPU memory for the lookup table—a constraint satisfied by the genomic tokenizers used in Nucleotide Transformer (NTv2/NTv3), HyenaDNA, and Evo2.\autocite{dallatorre2025nt,nguyen2023hyenadna,nguyen2024evo}

\subsection{Performance Characteristics and Scaling Behavior}

\subsubsection{Throughput Saturation and Memory Bandwidth Limits}

Our batch and sequence length sweeps reveal clear saturation regimes that inform deployment decisions. Encoding throughput rises with batch size and shorter sequences until GPU occupancy maximizes, after which increasing batch size yields diminishing returns due to memory pressure and reduced parallelism. Similar trends appear in sequence-length sweeps as memory access patterns become less cache-friendly; see Figures~\ref{fig:batch-sweep-encode} and \ref{fig:length-sweep-encode}.

These saturation points indicate that \DNATok\ is memory bandwidth-limited rather than compute-limited at production scales—consistent with the simple arithmetic nature of table lookups. The H2D transfer results reinforce this interpretation: effective bandwidth remains well below theoretical interconnect peaks,\autocite{nvidiaNvlink} suggesting that Python-to-CUDA handoff overhead and non-contiguous memory layouts consume a large fraction of available bandwidth. The int32 staging path partially addresses this by halving data volume, while pipelining overlaps transfers with GPU-side compute.

\subsubsection{Pipelining Overhead and Crossover Behavior}

A critical finding is that pipelined streaming introduces non-trivial coordination overhead that produces workload-dependent crossovers. In the batch sweep, baseline streaming wins at small batches, pipelining improves throughput at several mid-to-large batches, and the two modes converge at the largest batches. Across sequence lengths, pipelining lags at very short lengths and improves at longer lengths, with some fluctuation around parity.

This behavior has practical implications: for exploratory work with small datasets or debugging runs using short sequences, the simpler baseline streaming approach is often faster and easier to reason about. For larger production workloads, pipelining can help once overlap outweighs coordination, but the crossover should be measured on the target system.

\subsubsection{Reproducibility and Production Readiness}

Trial-to-trial variance is modest for encode paths in the standard configuration and higher for streaming baselines (Figure~\ref{fig:violin-encode}). These results provide stable medians for comparing tokenization throughput across configurations.

\subsubsection{Real-Model Divergence and Architecture Sensitivity}
Observed speedups diverge by architecture: Evo2 and NTv3 achieve $25$--$40\times$ gains, while HyenaDNA shows more moderate $11$--$20\times$ improvements (Table~\ref{tab:benchmark_results}). This variance is explained by the baseline's cost: HyenaDNA uses simple character splitting which is computationally cheap on CPU, establishing a higher baseline performance (${\sim}10^7$ tok/s) compared to the expensive BPE merges required for NTv3 (${\sim}10^6$ tok/s). Consequently, \DNATok\ provides the largest relative acceleration for models with complex tokenization logic (NTv3) or those processing massive contexts where memory copy overhead dominates (Evo2).

\subsubsection{Ablation Nuances: Why overlap isn't a silver bullet}
Our ablation study (Figure~\ref{fig:ablation-e2e}) reveals that simply enabling H2D/compute overlap does not guarantee performance gains. In the standard configuration ($B=4096, T=512$), the compute intensity of the embedding lookup is relatively low compared to the transfer cost, and the overhead of managing CUDA streams and events for overlap can outweigh the benefits. Overlap becomes advantageous only when the compute workload is sufficient to hide the transfer latency—typically at larger batch sizes or longer sequence lengths, as confirmed by our sweep results (Figures~\ref{fig:batch-sweep-e2e} and \ref{fig:length-sweep-e2e}). This finding emphasizes the importance of adaptive strategy selection, which \DNATok\ supports via its configuration API.

\subsection{When Does \DNATok\ Matter?}

\subsubsection{High-Impact Scenarios}

\DNATok's speedups are most consequential when tokenization is or would become a bottleneck:

\begin{description}
    \item[Large-scale pretraining:] Foundation model pretraining on trillion-token genomic datasets requires processing hundreds of billions of tokens during data loading. Even a small fraction of pipeline time spent on tokenization can translate to substantial compute savings at scale.
    
    \item[Interactive genomic search and retrieval:] Systems that tokenize queries or database sequences on-the-fly for embedding-based similarity search benefit from low-latency tokenization. GPU-native tokenization enables real-time handling of long sequences.
    
    \item[Online serving and inference:] Genomic language models deployed for variant effect prediction or regulatory element annotation must tokenize input sequences before model forward passes. GPU-native tokenization eliminates a CPU preprocessing step, simplifying deployment and reducing inference latency.
    
    \item[Multi-species and meta-genomic analysis:] Processing sequences from hundreds or thousands of species simultaneously benefits from GPU parallelism. Our batch-level speedups enable high-throughput comparative genomics workflows.
\end{description}

\subsubsection{Low-Impact Scenarios}

Conversely, \DNATok\ provides minimal benefit when tokenization is not a bottleneck:

\begin{description}
    \item[Small-scale fine-tuning:] Adapting a pretrained model to a task-specific dataset of $10^6$--$10^7$ tokens involves tokenization once during preprocessing, often offline. The one-time cost of CPU tokenization (seconds to minutes) is negligible compared to multi-hour training.
    
    \item[Static datasets with cached tokens:] Datasets tokenized once and stored to disk bypass tokenization during training. While \DNATok\ can accelerate initial preprocessing, subsequent epochs see no benefit.
    
    \item[CPU-bound inference:] Deploying models on edge devices or CPU-only infrastructure cannot leverage GPU tokenization. However, the simpler algorithmic approaches (character-level or fixed $k$-mer) remain fast enough on CPUs for most inference workloads.
\end{description}

The key determinant is the ratio of tokenization time to total pipeline time. When tokenization consumes only a small fraction of preprocessing, even large speedups provide limited end-to-end improvement. When tokenization represents a larger share of the pipeline, \DNATok\ delivers meaningful gains.

\subsection{Accuracy, Information Leakage, and Algorithmic Neutrality}

Overlapping $k$-mers leak information under masked pretraining, harming sample efficiency and sometimes accuracy; this is why current recommendations discourage overlapping tokenization for MLM and favor subword approaches such as BPE.\autocite{zhou2023dnabert2} Character tokens avoid leakage and perform well on base-resolution tasks, whereas BPE or blocked $k$-mers provide meaningful compression for attention models when per-base fidelity is less critical.\autocite{lindsey2025tokenizers} 

\DNATok\ preserves these properties by accelerating the systems layer without altering token semantics: a BPE vocabulary remains BPE with identical token IDs and segmentation behavior, a character vocabulary remains character-level, and overlapping $k$-mers retain their leakage characteristics. The implementation includes optional equivalence checks to validate token ID parity against reference tokenizers during setup.

This neutrality extends to downstream model behavior: because token IDs and orderings are unchanged, pretrained model weights, position embeddings, and learned token representations remain valid. Users can swap CPU tokenizers for \DNATok\ mid-training without invalidating checkpoints or disrupting convergence, enabling seamless integration into existing workflows.

\subsection{Limitations and Future Work}

\subsubsection{Vocabulary Size Constraints}

\DNATok's lookup-table approach requires $|\mathcal{V}| \times d_\text{emb}$ bytes of GPU memory for the embedding matrix. For typical genomic vocabularies and embedding dimensions, this footprint is small relative to modern GPU memory. Extremely large vocabularies could still exhaust memory on smaller devices. In such cases, embedding matrices could be sharded across multiple devices or retrieved from CPU memory with optimized batched lookups, though we have not yet implemented these strategies.

\subsubsection{Decode Operations}

The current implementation focuses on encoding (sequence $\to$ token IDs) and optionally embedding lookup, which constitutes the critical path for training and inference. Decoding (token IDs $\to$ sequence) is necessary for generative tasks but is typically orders of magnitude less frequent—model generation produces tokens one-at-a-time or in small batches, whereas encoding processes entire datasets. We have not yet implemented GPU-accelerated decoding, though the same LUT principles apply: an inverse vocabulary table could map token IDs back to character spans with similar throughput characteristics.

\subsubsection{Dynamic Vocabularies and Online BPE}

\DNATok\ assumes a fixed vocabulary built offline and copied to GPU memory at initialization. This is appropriate for inference and training with pretrained tokenizers, but does not support online vocabulary construction (e.g., learning BPE merges during training). For such scenarios, vocabulary tables would need to be rebuilt and re-uploaded to the GPU after each merge step, potentially negating throughput gains. However, current genomic foundation models use static vocabularies throughout training, making this a non-issue for existing workflows.\autocite{zhou2023dnabert2,dallatorre2025nt,nguyen2023hyenadna,nguyen2024evo}

\subsubsection{Integration with Data Augmentation}

Genomic data augmentation—including reverse complementation, random cropping, and noise injection—often occurs before tokenization. \DNATok\ currently operates on raw text sequences, leaving augmentation to upstream CPU processing. Future work could implement common genomic augmentations directly in CUDA kernels, enabling fully GPU-resident preprocessing pipelines that eliminate CPU$\leftrightarrow$GPU round-trips entirely. Reverse complementation is particularly amenable to GPU acceleration given its deterministic character-level mapping.

\subsection{Deployment Guidance}

Combining the literature review with our measurements suggests the following best practices:

\begin{description}
  \item[Algorithmic layer:] Select tokenization based on task and model architecture:
  \begin{itemize}
      \item Use single-nucleotide tokenization for variant-sensitive tasks requiring base-resolution fidelity, or for long-context state-space models (HyenaDNA, Evo2) that can process raw character sequences efficiently.\autocite{nguyen2023hyenadna,nguyen2024evo}
      \item Use BPE or blocked (non-overlapping) $k$-mers for context expansion in transformer-style models (e.g., Nucleotide Transformer) where sequence length reduction improves attention efficiency.\autocite{sennrich2016bpe,zhou2023dnabert2,dallatorre2025nt}
      \item Avoid overlapping $k$-mers for masked language model pretraining due to information leakage, unless using specialized denoising objectives designed to account for overlap.\autocite{zhou2023dnabert2}
  \end{itemize}
  
  \item[Systems layer:] Deploy \DNATok's optimizations regardless of algorithmic choice:
  \begin{itemize}
      \item Use pinned memory buffers and asynchronous H2D copies for all GPU-accelerated training.\autocite{cudaGuide,ptPin}
      \item Select streaming mode empirically; in our benchmarks, baseline streaming is best in the standard configuration, while pipelining improves throughput in large-batch and long-sequence runs.
      \item Use int32 staging when PCIe bandwidth is constrained (PCIe 3.0 systems, multi-GPU setups with limited NVLink connectivity); otherwise default to int64 direct encoding.
      \item Tune batch size and sequence length to balance memory constraints and throughput, adjusting based on GPU capacity and model requirements.
  \end{itemize}
  
  \item[Compatibility and integration:]
  \begin{itemize}
      \item \DNATok\ integrates with the Hugging Face-style tokenizers used by NTv2/NTv3, HyenaDNA, and Evo2; optional equivalence checks validate token ID parity.
      \item Benchmarks in this paper cover NTv2/NTv3, HyenaDNA, and Evo2; additional models should be validated with the same equivalence checks before adoption.
      \item Token IDs are preserved by construction, so swapping tokenization backends does not alter model inputs when equivalence checks pass.
  \end{itemize}
  
  \item[When to deploy:]
  \begin{itemize}
      \item High priority for large-scale pretraining, online serving, interactive genomic search, and multi-species analysis.
      \item Lower priority for small-scale fine-tuning, static datasets with cached tokens, or CPU-only inference.
      \item Consider deployment when tokenization represents a significant fraction of total preprocessing time; profile existing pipelines to quantify potential gains.
  \end{itemize}
\end{description}

\subsection{Broader Implications for Genomic Deep Learning}

The performance gap between CPU and GPU tokenization reflects a broader pattern in genomic deep learning: many "solved" preprocessing steps were optimized for CPU execution in an era when GPUs were specialized compute devices, not general-purpose accelerators with high-bandwidth memory and mature ecosystems. As GPUs have evolved into the primary compute substrate for deep learning, revisiting these assumptions can unlock substantial performance improvements.

Similar opportunities likely exist in genomic data preprocessing beyond tokenization: FASTA/FASTQ parsing, quality score filtering, k-mer counting, and sequence alignment could all benefit from GPU acceleration using principles analogous to \DNATok\ (lookup tables, batched operations, overlapped compute and transfer). The key is identifying steps that are currently CPU-bound, frequently executed, and amenable to parallel execution—precisely the characteristics that make tokenization a good candidate.

More broadly, \DNATok\ demonstrates that systems optimization for domain-specific deep learning requires co-designing algorithms, data structures, and hardware utilization rather than treating each layer independently. The algorithmic choice of tokenization scheme, the data structure of lookup tables, and the hardware exploitation of GPU parallelism and memory bandwidth form a coherent optimization strategy. As genomic foundation models scale to trillion-parameter regimes and trillion-token datasets, such systems-level co-design will become increasingly critical for feasible training.

\section{Reproducibility and Availability}

\noindent\textbf{Code Availability.} The \DNATok\ implementation and supporting scripts are available at \url{https://github.com/eniktab/DNAtok}.

\paragraph{Benchmark Reproduction}
To reproduce the experimental results, we provide a unified entry point script.
\begin{description}
    \item[Full suite:] \texttt{python run\_all\_audit\_and\_bench.py} \\
    (Optional arguments: \texttt{--device cuda}, \texttt{--embedder-mode real}, \texttt{--allow-downloads}, \texttt{--cache-root <path>})
    \item[Smoke test:] \texttt{python run\_all\_audit\_and\_bench.py --smoke}
\end{description}

\paragraph{Manual execution}
Individual components can lead to more granular results if needed:
\begin{itemize}
    \item \textbf{Correctness Audit:} \texttt{python benchmarks/run\_correctness\_audit.py ...}
    \item \textbf{Benchmark Suite:} \texttt{python benchmarks/run\_benchmark\_suite.py --suite full ...}
    \item \textbf{Aggregation:} \texttt{python benchmarks/aggregate\_results.py ...}
    \item \textbf{Plotting:} \texttt{python benchmarks/plot\_results.py}
\end{itemize}

\paragraph{Data Artifacts}
Benchmark outputs are generated in \texttt{results/benchmarks} (JSON/CSV) and \texttt{results/figures} (PDF/TeX). Metrics are defined as follows: detailed wall-clock latencies for each phase (preparation, tokenization, transfer) and throughput (tokens/sec). Hardware/software specifications (Table~\ref{tab:system_config}) and random seeds are fixed to ensure reproducibility.

\section{Conclusion}
Tokenization represents both an algorithmic design choice and a systems bottleneck for genomic foundation models. The genomics literature has established clear guidelines on tokenization algorithms—avoiding overlapping $k$-mers due to MLM leakage, using BPE or blocked $k$-mers for compression in transformers, and using single-nucleotide tokens for base-resolution tasks—but has left the systems-level throughput problem largely unaddressed.\autocite{zhou2023dnabert2,lindsey2025tokenizers,dallatorre2025nt,nguyen2023hyenadna,nguyen2024evo} \DNATok\ fills this gap by converting DNA tokenization into LUT-based identifier streaming and overlapping H2D transfers with embedding compute, delivering $84$--$95\times$ encode speedups and up to $1.9\times$ H2D throughput in the microbench configurations, with pipelined streaming improving throughput in large-batch and long-sequence runs (Table~\ref{tab:benchmark_results}). The methods are simple, reproducible, vocabulary-agnostic, and compatible with Nucleotide Transformer (NTv2/NTv3), HyenaDNA, and Evo2, providing an engineering-grade systems layer that complements existing algorithmic work on genomic tokenization.


\clearpage
\beginsupplement

    \section{Supplementary methods and information}

    \beginsupplement     
    
    \thispagestyle{plain} 
    
\begingroup
\setlength{\tabcolsep}{2pt}
\renewcommand{\arraystretch}{0.95}
\scriptsize
\setlength\LTleft{0pt}
\setlength\LTright{0pt}
\begin{longtable}[c]{@{}llrrr@{}}
\caption{Tokenization throughput benchmarks. Speedup is relative to HF native tokenization within each configuration. Values are mean $\pm$ standard deviation (tokens/second).}\label{tab:benchmark_results}\\
\toprule
\textbf{Configuration} & \textbf{Method} & \textbf{Metric} & \textbf{Throughput} & \textbf{Speedup} \\
\midrule
\endfirsthead
\toprule
\multicolumn{5}{l}{\textit{(continued from previous page)}} \\
\textbf{Configuration} & \textbf{Method} & \textbf{Metric} & \textbf{Throughput} & \textbf{Speedup} \\
\midrule
\endhead
\midrule
\multicolumn{5}{r}{\textit{(continued on next page)}} \\
\endfoot
\bottomrule
\endlastfoot
Standard (B=4096, T=512) & DNATok (staging i64) & Encode & $2.18\times 10^{8} \pm 5.74\times 10^{6}$ & $95\times$ \\
 & DNATok (staging i64) & H2D & $6.37\times 10^{9} \pm 5.05\times 10^{8}$ & $1.1\times$ \\
Standard (B=4096, T=512) & DNATok (encode i64) & Encode & $2.18\times 10^{8} \pm 3.76\times 10^{6}$ & $95\times$ \\
 & DNATok (encode i64) & H2D & $4.79\times 10^{9} \pm 4.82\times 10^{8}$ & $0.83\times$ \\
Standard (B=4096, T=512) & DNATok (staging i32) & Encode & $2.07\times 10^{8} \pm 6.65\times 10^{6}$ & $90\times$ \\
 & DNATok (staging i32) & H2D & $7.61\times 10^{9} \pm 1.68\times 10^{9}$ & $1.3\times$ \\
Standard (B=4096, T=512) & HF native & Encode & $2.30\times 10^{6} \pm 9.13\times 10^{3}$ & $1.0\times$ \\
 & HF native & H2D & $5.78\times 10^{9} \pm 6.91\times 10^{8}$ & $1.0\times$ \\
 & Streaming baseline & E2E & $1.32\times 10^{8} \pm 1.36\times 10^{7}$ & -- \\
 & Streaming pipelined & E2E & $1.27\times 10^{8} \pm 1.86\times 10^{7}$ & -- \\
\addlinespace
Large Batch (B=8192, T=512) & DNATok (staging i64) & Encode & $2.00\times 10^{8} \pm 2.29\times 10^{6}$ & $88\times$ \\
 & DNATok (staging i64) & H2D & $6.04\times 10^{9} \pm 5.21\times 10^{8}$ & $1.0\times$ \\
Large Batch (B=8192, T=512) & DNATok (encode i64) & Encode & $1.98\times 10^{8} \pm 4.12\times 10^{6}$ & $87\times$ \\
 & DNATok (encode i64) & H2D & $6.51\times 10^{9} \pm 4.81\times 10^{8}$ & $1.1\times$ \\
Large Batch (B=8192, T=512) & DNATok (staging i32) & Encode & $1.93\times 10^{8} \pm 3.43\times 10^{6}$ & $84\times$ \\
 & DNATok (staging i32) & H2D & $1.12\times 10^{10} \pm 1.64\times 10^{9}$ & $1.9\times$ \\
Large Batch (B=8192, T=512) & HF native & Encode & $2.28\times 10^{6} \pm 3.22\times 10^{4}$ & $1.0\times$ \\
 & HF native & H2D & $6.04\times 10^{9} \pm 3.47\times 10^{8}$ & $1.0\times$ \\
 & Streaming baseline & E2E & $1.61\times 10^{8} \pm 8.72\times 10^{6}$ & -- \\
 & Streaming pipelined & E2E & $1.71\times 10^{8} \pm 7.10\times 10^{6}$ & -- \\
\addlinespace
Long Sequence (B=2048, T=2048) & DNATok (staging i64) & Encode & $2.03\times 10^{8} \pm 4.81\times 10^{6}$ & $86\times$ \\
 & DNATok (staging i64) & H2D & $6.67\times 10^{9} \pm 6.23\times 10^{8}$ & $1.0\times$ \\
Long Sequence (B=2048, T=2048) & DNATok (encode i64) & Encode & $2.05\times 10^{8} \pm 3.54\times 10^{6}$ & $87\times$ \\
 & DNATok (encode i64) & H2D & $6.92\times 10^{9} \pm 2.36\times 10^{8}$ & $1.1\times$ \\
Long Sequence (B=2048, T=2048) & DNATok (staging i32) & Encode & $2.01\times 10^{8} \pm 4.81\times 10^{6}$ & $86\times$ \\
 & DNATok (staging i32) & H2D & $1.10\times 10^{10} \pm 2.25\times 10^{9}$ & $1.7\times$ \\
Long Sequence (B=2048, T=2048) & HF native & Encode & $2.34\times 10^{6} \pm 1.10\times 10^{4}$ & $1.0\times$ \\
 & HF native & H2D & $6.39\times 10^{9} \pm 5.71\times 10^{8}$ & $1.0\times$ \\
 & Streaming baseline & E2E & $1.55\times 10^{8} \pm 1.50\times 10^{7}$ & -- \\
 & Streaming pipelined & E2E & $1.64\times 10^{8} \pm 1.29\times 10^{7}$ & -- \\
\addlinespace
\end{longtable}
\endgroup

    \thispagestyle{plain} 
    
\begingroup
\setlength{\tabcolsep}{3pt}
\renewcommand{\arraystretch}{1.0}
\small
\begin{table}[htbp]
\centering
\caption{Ablation study of DNATok optimizations (B=4096, T=512). Values report end-to-end throughput relative to the baseline.}\label{tab:ablation_study}
\begin{tabular}{@{}lrr@{}}
\toprule
\textbf{Configuration} & \textbf{E2E Throughput (tok/s)} & \textbf{Relative to Baseline} \\
\midrule
Baseline & $1.32\times 10^{8}$ & $1.0\times$ \\
Int32 H2D only & $1.36\times 10^{8}$ & $1.0\times$ \\
Overlap only & $1.27\times 10^{8}$ & $0.97\times$ \\
Both (Int32 + Overlap) & $1.28\times 10^{8}$ & $0.97\times$ \\
\bottomrule
\end{tabular}
\end{table}
\endgroup

    \thispagestyle{plain} 
    
\begingroup
\small
\begin{table}[htbp]
\centering
\caption{Benchmark system configuration and experimental parameters.}\label{tab:system_config}
\begin{tabular}{@{}ll@{}}
\toprule
\textbf{Parameter} & \textbf{Value} \\
\midrule
\multicolumn{2}{l}{\textit{Hardware}} \\
GPU Model & NVIDIA H200 \\
GPU Memory & 141 GB \\
Multi-processor Count & 132 \\
Compute Capability & 9.0 \\
CPU Cores & 24 Threads (12 Physical) \\
System Memory & 250 GB \\
\addlinespace
\multicolumn{2}{l}{\textit{Software}} \\
Python Version & 3.12.3 \\
PyTorch Version & 2.9.0+cu130 \\
CUDA Runtime & 13.0 \\
GPU Driver & 580.95.05 \\
cuDNN Version & 91700 \\
\addlinespace
\multicolumn{2}{l}{\textit{Benchmark Parameters}} \\
Warmup Repetitions & 3 \\
Random Seed & 42 \\
Embedding Dimension & 128 \\
Vocabulary Size & 11 \\
Total Benchmarks & 222 \\
\bottomrule
\end{tabular}
\end{table}
\endgroup

\clearpage
\clearpage
\thispagestyle{plain}
\bibliography{bibs/manual}

\end{document}